\documentclass[eqsecnum,floats,aps,nofootinbib,showpacs]{revtex4}
\usepackage{latexsym}
\usepackage{amssymb}
\usepackage{hyperref}

\renewcommand{\d}{\mathrm{d}}
\renewcommand{\l}{\left(}
\renewcommand{\r}{\right)}
\usepackage{amsmath, amsthm, amssymb}

\def\be{\begin{equation}}
\def\ee{\end{equation}}
\def\beq{\begin{equation*}}
\def\eeq{\end{equation*}}
\def\ba{\begin{aligned}}
\def\ea{\end{aligned}}

\def\ov{\overline}
\def\w{\wedge}

\begin{document}
\title{Einstein--Cartan gravity with Holst term and fermions}

\author {Marcin Ka\'zmierczak}
\email{marcin.kazmierczak@fuw.edu.pl}
\affiliation{Institute of Theoretical Physics, University of Warsaw,
  Ho\.{z}a 69, 00-681 Warszawa, Poland}

\begin{abstract}
We investigate the consequences of the ambiguity of minimal coupling
procedure for Einstein--Cartan gravity with Holst term and fermions. A
new insight is provided into the nature and physical relevance of coupling
procedures considered hitherto in the context of
Ashtekar--Barbero--Immirzi formalism with fermions. The issue
of physical effects of the Immirzi parameter in semi--classical theory is
reinvestigated. We argue that the conclusive answer to the question of
its measurability will not be possible until the more fundamental problem
of nonuniqueness of gravity--induced fermion interaction in
Einstein--Cartan theory is solved.
\end{abstract}
 
\maketitle

\section{Introduction}
\label{section1}
In the last few years one can observe an increase of interest in
coupling fermionic matter to Einstein-Cartan (EC) gravity modified by
the presence of a new term in the gravitational action, the so called Holst
term \cite{PR,FMT,Ran,Merc,BD,Alex}. Its inclusion simplifies the structure of constraints of the
theory and enables
nonperturbative canonical quantisation
of gravitational field in the lines of Loop Quantum Gravity (LQG)
program. The history of this term can be traced back to the
introduction by Ashtekar of new varibles for general relativity (GR)
\cite{A1}\cite{A2}, reducing
constraints of the theory to the
polynomial form. In order to avoid difficulties
concerning reality conditions, necessary in Ashtekar complex
approach, Barbero \cite{Bar} proposed a real alternative. The
relation between these two approaches was then clarified by
Immirzi \cite{Imm} and Holst \cite{Hol}. It appeared that the addition
of a new term to the gravitational action allows for a unified treatment
of both approaches. The real parameter $\beta$ whose inverse precedes the new
term is called an {\it Immirzi parameter}. In the covariant Holst
approach, the Ashtekar and Barbero formulation can be recovered for
$\beta=\pm i$ and $\beta=\pm 1$, respectively. However, it appeared
that one can proceed in the program of canonical quantisation for
$\beta$ assuming a generic real value \cite{AL}. The parameter proved
to play an important role in the quantum theory of geometry by
entering the spectra of area and volume operators \cite{RT}. 

The Holst modification of the standard Palatini action does not change classical
field equations, so long as matter with nonvanishing spin distribution
is not included. However, the first order formalism employed by LQG leads in a
natural way to EC gravity, rather than standard GR. Hence, the
space--time manifold acquires a non--vanishing
torsion when the spin of matter is not negligible. This makes the Holst term modify field
equations. In \cite{PR}, fermionic matter was considered, modeled
by the standard Dirac Lagrangian. It was concluded that the Immirzi
parameter can be observable
in principle, even if quantum aspects of gravitational field are not
taken into account. The coupling constant in front of the well known
term describing point fermion interaction of EC gravity
\cite{HD,Ker,Rumpf} appeared to
involve Immirzi parameter. The possibility of establishing experimental
bounds on the parameter without invoking the complete theory of
quantum geometry would be of
great importance, since it seems there is a long way for LQG to develop before
definite experimental predictions can be extracted. One of the
successes of LQG has been the derivation of the formula for black hole entropy, which
appeared to involve the Immirzi parameter. This allowed for the
establishing of theoretical bounds on possible values of the Immirzi
parameter by black hole entropy calculations and comparison with the
Bekenstein--Hawking formula \cite{DL}. A precise value of
the parameter was given shortly after \cite{Meiss}.
If we were also able to find experimental limits on this value,
as it was hoped in \cite{PR}, a comparison of these two results would
provide the first test of physical relevance of LQG approach.

The minimal coupling of fermions to gravity was employed in
\cite{PR}. The necessity of that coupling scheme and hence all physical
predictions resulting from \cite{PR} were questioned shortly
after. Two different one--parameter families of possible non--minimal
couplings were suggested in \cite{FMT} and \cite{Merc}. The
family considered in \cite{FMT}, which was claimed general,
led authors to the conclusion that the minimally coupled theory is
parity invariant, whereas the non--minimally coupled is not. This conclusion was
based on the analysis of transformation properties of the effective
action, obtained by establishing the connection between torsion and
matter and inserting the result back to the initial action so that
torsion was effectively eliminated from the theory. In fact,
the minimally coupled theory is also not parity invariant, although its
effective action is. To see this, one should take into account the
transformation properties of
equations connecting torsion with matter fields. Parity
breaking in the case of minimal coupling was first observed in
\cite{Merc}, but it was misinterpreted as an indication of
internal inconsistency of a theory. After criticizing the minimal
procedure, a one--parameter family of non--minimal couplings was
proposed in \cite{Merc}, apparently different from the one considered in \cite{FMT} (in
contradiction with the statements of \cite{FMT} concerning generality). Parity
violation was avoided by an appropriate choice of the parameter of
the non--minimal coupling, suitably adopted to the value of the Immirzi
parameter. After this choice was made, the theory reduced to the
one which is often viewed as usual EC gravity with fermions (as
explained in \cite{Kazm}, the hitherto accounts on EC gravity with fermions
were farily incomplete). Finally, in \cite{Alex} there was
considered the four--parameter family of non--minimal couplings
generalising the two families of \cite{FMT} and \cite{Merc}. It has
been concluded that the Immirzi parameter is not measurable, as it can
be ``hidden'' in the coupling parameters. It has not been taken into
account that the possibility of measuring the Holst term
induced physical effects might be regained, if we were able to
decide on theoretical grounds which coupling procedure is physically
appropriate.

The leading idea of this paper is that the minimal coupling scheme is
the one that has been historically successful in constructing models, which could
withstand the rigors of experimental testing whenever such tests were
feasible. The experimental successes of the
standard model of particle physics and general theory of
relativity seem to support the minimal approach. Indeed, the Yang--Mills theories, which constitute the
formal basis for the standard model, employ minimal coupling on
the fundamental level. The necessity of using non--minimal couplings
when describing effectively composed objects does not hold much relevance
as long as we aim to incorporate elementary point--like fermions (quarks and
leptons) into the theory of gravity. As is well known, the EC
gravity can be formulated as a gauge theory of Yang-Mills type for the
Poincar{\'e} group \cite{Kib1,HHKN,T1,GN}. Hence, one could hope that the application of
minimal coupling would lead to the physically relevant model in this case as
well. According to this viewpoint, special importance should be
attached to the original analysis of \cite{PR},
rather than to the later analyses employing
non--minimal couplings. It is also important to stress that the conjecture of
\cite{Merc} that the Holst term should not produce any physical
effects in the classical theory is not justified. We cannot know that before a thorough
theoretical analysis is completed.

There is however an issue of fundamental importance that makes things
more involved. If torsion of space--time does not vanish, the standard
minimal coupling procedure (MCP) is not unique. Equivalent flat space Lagrangians (generating the same flat
space field equations) give rise to generically non--equivalent
curved theories. The problem has been discussed since the very beginning of
gauge formulation of gravity \cite{Kib1}. Apart from the search for the
criteria of choice of the most physical flat space Lagrangian, a more radical
solution was proposed by Saa in \cite{Saa1}\cite{Saa2}. The introduction
of a connection--compatible volume element, instead of a metric--compatible one, together with the
requirement for torsion trace to be derivable from a potential,
eliminated the ambiguity. The new procedure leads to interesting
effects, such as a propagating torsion or coupling gauge
fields to torsion without breaking gauge symmetry. Although Saa's idea
provides a very interesting solution to the problem, it results in
significant departures from standard GR, which are not certain to
withstand the confrontation with observable data \cite{BFY}\cite{FY} without
some assumptions of rather artificial nature, such as demanding
a priori that part of the torsion tensor vanish \cite{RMAS}.

The consequences of the above--mentioned ambiguity for EC theory
with fermions were investigated in \cite{Kazm}. It appears that what
was considered in \cite{PR,Merc} as standard EC theory with
fermions is only one possibility from the two--parameter family of
theories, this family being related to the 
freedom of divergence addition to the flat space Lagrangian density (see Section \ref{NU}). After this result is at hand, one can no longer acknowledge the
analysis of \cite{PR} as satisfying, as it corresponds to a particular
flat fermionic Lagrangian randomly selected from the infinity of
possibilities. In this paper we aim to exhaust the possibilities left
by MCP. Contrary to the approach presented in
\cite{PR}, we aim to limit the multiplicity of ``equivalent'' flat space Lagrangians by imposing reasonable restrictions on them,
rather than choosing a particular one without any justification.

An interesting result of the following paper is the observation that all the non--equivalent theories
obtained by different non--minimal coupling procedures of
\cite{FMT,Merc,Alex} can be interpreted as resulting from
the application of MCP to the suitably chosen fermionic flat
space Lagrangians (see Section \ref{NMC}). Hence, even if MCP is given the priority, we cannot claim those procedures to be worse
than the one of \cite{PR}. However, from this viewpoint, the problem
of choosing among them may be reduced to the choice of the most physically reasonable flat space
Lagrangian for fermions. Some suggestions concerning this choice are
given at the end of Section \ref{NMC}.

The paper is organized as follows: 
In Section \ref{NU} we explain the origins of ambiguity of MCP and impose some obvious restrictions on flat space
fermionic Lagrangians, which leaves us with a two--parameter family.
In Section \ref{NMC} we show that all the coupling procedures discussed so
far in the context of fermions in
Ashtekar--Barbero--Immirzi formalism can be realised as minimal
couplings for appropriate flat space Lagrangians for fermions. We
comment on their physical relevance. We also discuss a truly
non--minimal procedure, which cannot be reinterpreted in this manner.   
In Section \ref{ECgravity} we briefly recall the
formalism of EC theory with Holst term and rederive the effective
action taking into account the freedom of addition of a divergence to
the flat space matter Lagrangian. We also comment on the possibility
of detecting the physical effects produced by the Immirzi parameter. 
In Section \ref{conc} we draw conclusions.

\section{Nonuniqueness of minimal coupling procedure}\label{NU}
A classical field theory in flat Minkowski space is defined by the
action functional
\beq
S=\int \mathfrak{L} \ ,
\eeq
where $\mathcal{L}$ is a Lagrangian density and
$\mathfrak{L}=\mathcal{L}\, \d^4 x=\mathcal{L}\,\d x^0\wedge\d
x^1\wedge\d x^2\wedge\d x^3$ a Lagrangian four--form. 
It is well known that the addition of a divergence of a vector field $V$ to
$\mathcal{L}$ changes $\mathfrak{L}$ by a differential
\be\label{differential}
\partial_{\mu}V^{\mu}\, \d^4 x=L_V\,\d^4 x=\d (V \lrcorner \, \d^4 x) \ ,
\ee
where $L$ denotes Lie derivative and $\lrcorner$ the internal
product. Thus, such a transformation does not change field equations
generated by $S$. In order to proceed from Minkowski space to the general
Riemann--Cartan (RC) manifold with metric $g_{\mu\nu}$ and the metric
compatible connection
$\nabla$ (not--necessarily torsion free)\footnote{For more general
  considerations concerning not--necessarily metric connections see \cite{RMAS}.}, we can apply MCP
\be\label{MCP}
\int \mathcal{L}(\phi,\partial_{\mu}\phi,\dots )\,\d^4 x \
\longrightarrow \ \int \mathcal{L}(\phi,\nabla_{\mu}\phi,\dots
)\,\epsilon \ ,
\ee
where $\epsilon =\sqrt{-g}\,\d^4 x$ is a volume--form, $g$ being the
determinant of a matrix of components $g_{\mu\nu}$ of the metric
tensor in the basis $\partial_{\mu}$, and $\phi$ represents fields of
the theory. Dots in (\ref{MCP}) correspond
to the possibility of $\mathcal{L}$ to depend on higher derivatives of
fields. Had we used the modified flat space Lagrangian
$\mathcal{L}+\partial_{\mu}V^{\mu}$, we would have obtained a different
Lagrangian four--form on the RC manifold, the difference being
\be\label{TV}
\nabla_{\mu}V^{\mu}\epsilon={\stackrel{\circ}{\nabla}}_{\mu}V^{\mu}\epsilon-T_{\mu}V^{\mu}\epsilon
\ ,
\ee
where $\stackrel{\circ}{\nabla}$ is the torsion free Levi--Civita connection and
$T_{\mu}={T^{\nu}}_{\mu\nu}$ the torsion trace vector. The first term
in (\ref{TV}) is a differential,
${\stackrel{\circ}{\nabla }}_{\mu}V^{\mu}\epsilon=\d (V \lrcorner \,
\epsilon )$,
whereas the second is not. Hence, the equivalent flat Lagrangians yield
non--equivalent theories on RC space. 

One could hope differential forms formalism would fix the problem and
argue that the last expression of
(\ref{differential}), rather than the first, should be adopted to curved space. Then, $\d (V \lrcorner \,
\d^4 x)$ would transform into $\d (V \lrcorner \, \epsilon)$, which is
again a differential. However, this is not a good solution, since decomposition of a given Lagrangian four--form 
$\mathcal{L}_1\,\d^4 x$ to the sum of another Lagrangian four--form $\mathcal{L}_2\,\d^4 x$ and
the term $\d (V\lrcorner \,\d^4 x)$ is by no means unique. We should
rather use the identity $\d (V \lrcorner \, \d^4 x)=-\l{\star \d x_{\mu}}\r
\wedge \d V^{\mu}$, where $\star$ is a hodge star (see Section \ref{App1}), and minimally couple gravity by the passage
$\d V^{\mu}\longrightarrow DV^{\mu}=\d V^{\mu}+{\omega^{\mu}}_{\nu}
V^{\nu}$ (where ${\omega^{\mu}}_{\nu}$ are connection one--forms) and by the
change of a hodge star of flat Minkowski metric to the one of curved
metric on the finall manifold, but this would give the result
identical to (\ref{TV}).

Let us now consider two Lagrangian densities differing by a divergence
of a vector field $V^{\mu}(\phi)$ (we wish $V^{\mu}$ not to depend
on derivatives of $\phi$ in order for both Lagrangians to depend on
first derivatives only)
\be\label{Lch}
\mathcal{L}-\mathcal{L}'=\partial_{\mu}V^{\mu}=\frac{\partial
  V^{\mu}}{\partial \phi}\partial_{\mu}\phi \ .
\ee
Here, $V$ is required to transform as a vector under proper Lorentz
transformations: if $\phi\rightarrow \phi'$ represents the
action of a relevant representation of a proper Lorentz group in the
space of fields, we have $V^{\mu}\l{\phi}\r\rightarrow
V^{\mu}\l{\phi'}\r={\Lambda^{\mu}}_{\nu}V^{\nu}\l{\phi}\r$. Hence,
$\partial_{\mu}V^{\mu}$ is a Lorentz scalar and $\mathcal{L}'$
is a Lorentz scalar (if $\mathcal{L}$ is). All Lagrangian densities considered by us are also
required to be real, which implies the reality of $V$.
 Let us then focus our attention on the Dirac field $\psi$. The
 requirement for Lagrangians to be real and quadratic in the fields suggests
 the following form of $V$
\be\label{VB}
V^{\mu}=\ov{\psi}B^{\mu}\psi \ ,
\ee
where the matrixes $B^{\mu}$ obey the reality condition
${B^{\mu}}^{\dagger}=\gamma^0 B^{\mu}\gamma^0$. Together with the
requirement of vector transformation properties under the action of
proper Lorenz group, it leads to $B^{\mu}=a\gamma^{\mu}+b\gamma^{\mu}\gamma^5$ for some real
numbers $a$ and $b$. Here $\gamma^{\mu}$ are the Dirac matrixes obeying 
$\gamma^{\mu}\gamma^{\nu}+\gamma^{\nu}\gamma^{\mu}=2\eta^{\mu\nu}$, 
$\gamma^5:=-i\gamma^0\gamma^1\gamma^2\gamma^3$ and
$\ov{\psi}:={\psi}^{\dagger}\gamma^0$, where ${\psi}^{\dagger}$ is
a Hermitian conjugation of a column matrix. Hence, 
\be\label{Vab}
V^{\mu}=aJ_{(V)}^{\mu}+bJ_{(A)}^{\mu} \ ,
\ee
where $J_{(V)}^{\mu}=\ov{\psi}\gamma^{\mu}\psi$,
$J_{(A)}^{\mu}=\ov{\psi}\gamma^{\mu}\gamma^5\psi$ denote Dirac vector and
axial current.

\section{Non--minimal couplings from a different perspective}\label{NMC}

\subsection{Apparent non--minimal couplings}\label{ANMC}

According to \cite{Alex}, the general non--minimally coupled fermion
action quadratic in fermionic field is an integral of a Lagrangian four--form
\be
\mathfrak{L}_F=\frac{i}{2}\left({\ov{\psi}\gamma^a (\zeta-i\xi
    \gamma^5)D_a\psi-\ov{D_a\psi}(\ov{\zeta}-i\ov{\xi}\gamma^5)\gamma^a\psi}\right)\,\epsilon \ ,
\ee
where bar means complex conjugation while acting on numbers and Dirac
conjugation while acting on spinors (see Section \ref{App1} for
the definition of $D_a$). In fact, this action can be
obtained via MCP from the flat space Lagrangian
\be\label{LF}
\mathcal{L}_F=\frac{i}{2}\left({\ov{\psi}\gamma^{\mu} (\zeta-i\xi
    \gamma^5)\partial_{\mu}\psi-\partial_{\mu}\ov{\psi}(\ov{\zeta}-i\ov{\xi}\gamma^5)\gamma^{\mu}\psi}\right) \ .
\ee
Following Alexandrov, we shall denote 
\be
\zeta=\eta+i\theta \ , \quad  \xi=\rho+i\tau \ , \qquad 
\eta, \theta, \rho, \tau \in \mathbb{R} \ .
\ee
Then we get
\be
\ba
\mathcal{L}_F=\,
\eta\frac{i}{2}\left({\ov{\psi}\gamma^{\mu}\partial_{\mu}\psi-\partial_{\mu}\ov{\psi}\gamma^{\mu}\psi}\right)-
\frac{\theta}{2}\partial_{\mu}(\ov{\psi}\gamma^{\mu}\psi)+
\frac{\rho}{2}\partial_{\mu}(\ov{\psi}\gamma^{\mu}\gamma^5\psi)+
\tau\frac{i}{2}\left({\ov{\psi}\gamma^{\mu}\gamma^5\partial_{\mu}\psi-\partial_{\mu}\ov{\psi}\gamma^{\mu}\gamma^5\psi}
\right) \ .
\ea
\ee
As we can now see, the first component is just the mass--free part of
the standard Dirac Lagrangian density
\be\label{LF0}
\mathcal{L}_{F0}=\frac{i}{2}\left({\ov{\psi}\gamma^{\mu}\partial_{\mu}\psi-\partial_{\mu}\ov{\psi}\gamma^{\mu}\psi}\right)
-m\ov{\psi}\psi \ ,
\ee
multiplied by a scaling constant $\eta$,
which could be set to one. The next two terms are divergences of vector and axial currents and they correspond to the families of
`non--minimal' couplings considered in \cite{FMT} and \cite{Merc},
respectively. What is the most interesting is the last term. It seems to have
been overlooked in \cite{FMT} and \cite{Merc} and brought to life in
\cite{Alex}. The Lagrangian obtained by setting $\theta=\rho=0$, which
can be conveniently rewritten in the form
\be\label{unph}
\mathcal{L}_1=\frac{i}{2}
\left[{ \ \ov{\psi}\l{\eta-\tau\gamma^5}\r\gamma^{\mu}\partial_{\mu}\psi-
\partial_{\mu}\ov{\psi}\l{\eta-\tau\gamma^5}\r\gamma^{\mu}\psi \ }\right] \ ,
\ee
generates the equation
\be
\l{\eta-\tau\gamma^5}\r i\gamma^{\mu}\partial_{\mu}\psi=0,
\ee
which is equivalent to the mass--free Dirac equation if $\eta^2\not=
\tau^2$ (under this condition the matrix $\eta-\tau\gamma^5$
is invertible).
At first, one could suppose that (\ref{unph}) might be generalised by
\be\label{LA}
\mathcal{L}_A=\frac{i}{2}
\left[{ \ \ov{\psi} A \gamma^{\mu}\partial_{\mu}\psi-
\partial_{\mu}\ov{\psi} A \gamma^{\mu}\psi \ }\right] ,
\ee
where $A$ is an invertible $4 \times  4$  matrix, but the reality
condition for (\ref{LA}) 
\be\label{real1}
\gamma^{\mu}\gamma^0 A^{\dagger}=A\gamma^{\mu}\gamma^0
\ee
implies that
$A=\eta-\tau\gamma^5$, where $\eta,\tau$ are real numbers, leaving us
with (\ref{unph}). In the following, we will use the letter $A$ just
to denote the matrix $\eta-\tau\gamma^5$, where the invertibility condition $\eta^2\not=
\tau^2$ should be understood to hold.

Let us investigate the possibility of adding a mass term to the Lagrangian (\ref{unph}). An obvious choice would be to add
\be\label{mass}
-m\ov{\psi}A\psi
\ee
to (\ref{unph}). Then the variation with respect to $\ov{\psi}$
would yield the massive Dirac equation. However, the Lagrangian would no
longer be  real, as the reality condition (\ref{real1}) does not guarantee
the reality of (\ref{mass}). The appropriate reality condition for
(\ref{mass}) is $\gamma^0A^{\dag}\gamma^0=A$, which will be fulfilled
if and only if $\tau=0$. Note that for a non--real Lagrangian it is
not guaranteed that the variation with respect to $\psi$ will give the
equation which is equivalent to the one obtained from variation with respect to
$\ov{\psi}$. Indeed, straightforward calculation of this variation
gives the equation 
\be
-i\partial_{\mu}\ov{\psi}A\gamma^{\mu}-m\ov{\psi}A=0.
\ee
After the Dirac conjugation is performed and the reality condition
(\ref{real1}) is applied, this equation can be rewritten as 
\be
iA\gamma^{\mu}\partial_{\mu}\psi-m\gamma^0A^{\dag}\gamma^0\psi=0.
\ee
It is now clear that it would not be equivalent to the massive Dirac
equation unless $\gamma^0A^{\dag}\gamma^0=A$ and hence $\tau=0$. One
could try the addition of a more general mass term $-m\ov{\psi}B\psi$,
where $B$ is a general $4\times 4$ matrix obeying the reality
condition $\gamma^0B^{\dag}\gamma^0=B$. Then the variation with
respect to $\ov{\psi}$ and $\psi$ would yield the same equation. But
this equation would not be equivalent to the Dirac one, unless $B=A$,
which leads again to $\tau=0$.

We have argued that it seems impossible to find an appropriate mass term
for the Lagrangian (\ref{unph}) for $\tau \not= 0$. All fermions that
have been detected in nature are massive. Although their masses are
not inserted {\it a'priori} into the Lagrangian, but rather arise as a
result of spontaneous symmetry breaking via Higgs mechanism, it is
still important that the initial Dirac Lagrangian allow for the
addition of massive term in a consistent way. Otherwise the Higgs
mechanism could not yield an expected result. This is why we
will not take into account the subfamily of (\ref{LF}) corresponding to
$\tau\not= 0$ in further considerations. We are therefore left with the two--parameter family of Lagrangians (we set $\eta=1$
to exclude also scaling) that can be constructed from (\ref{LF0}) by
the addition of a divergence of a linear combination of axial and vector
Dirac currents. As shown in the previous section, this is all the
freedom we can gain by adding a divergence to the Lagrangian density, under the
requirements for it to be real, second order in field powers, first
order in derivatives and invariant under the proper Lorenz
transformations. The consequences of this freedom will be exploited in Subsection \ref{effective}. In the following, we will use the parameters
$a=-\theta/2$, $b=\rho/2$ introduced in
(\ref{Vab}). It seems very difficult to reduce the remaining
freedom on theoretical grounds. In the case of EC theory without Holst modification, one
could demand the Lagrangian density to be parity invariant, which
would correspond to $b=0$. However, if we allow for the addition of Holst
term to the standard Palatini term of gravitational action, which
clearly behaves differently under parity transformation, there is no
reason for excluding pseudo--scalar term corresponding to $b\not= 0$
from the flat Lagrangian density.

\subsection{Genuine non--minimal couplings}\label{TNMC}
All the ``non--minimal'' couplings of gravity to fermions considered so
far could be reinterpreted as minimal couplings for suitable
fermionic Lagrangians. Are there any truly non--minimal couplings which
cannot be viewed as minimal ones for any choice of the flat space Lagrangian? To answer this question, let us
first note that the Dirac Lagrangian four--form obtained from
(\ref{LF0}) via MCP
\[
\tilde{\mathfrak{L}}_{F0}=
-\dfrac{i}{2}\l{\star e_a}\r \wedge \l{ \ov{\psi}\gamma^a
  D\psi-\ov{D\psi}\gamma^a\psi}\r-m\ov{\psi}\psi\,\epsilon  
\]
decomposes as
\be\label{LFdecomp}
\tilde{\mathfrak{L}}_{F0}={\stackrel{\circ}{\tilde{\mathfrak{L}}}_{F0}}-\frac{1}{8}S_aJ^a_{(A)}\,\epsilon \ ,
\ee
where ${\stackrel{\circ}{\tilde{\mathfrak{L}}}_{F0}}$ is a part
determined by the Levi--Civita connection. Here $S_a=\epsilon_{abcd}T^{bcd}$, where
${T^a}_{bc}$ are components of the torsion tensor in an unholonomic
tetrad basis (in a holonomic frame $\partial_{\mu}$ the components are given by
${T^{\rho}}_{\mu\nu}=-{\Gamma^{\rho}}_{\mu\nu}+{\Gamma^{\rho}}_{\nu\mu}$,
where the connection coefficients are deffined by
$\nabla_{\partial_{\mu}}\partial_{\nu}={\Gamma^{\rho}}_{\nu\mu}\partial_{\rho}$).
This form of decomposition
suggests the non--minimal coupling
\be\label{GNC}
\mathfrak{L}_{F,nm}={\stackrel{\circ}{\tilde{\mathfrak{L}}}_{F0}}+\eta_1\,
S_aJ^a_{(A)}\,\epsilon \ ,
\ee
where $\eta_1\in\mathbb{R}$ is a coupling parameter. 
This family cannot be attained by MCP from any flat Lagrangian. In
fact, there does not exist a flat space Lagrangian which would produce
via MCP the Lagrangian four-form differing from (\ref{GNC}) by a
differential. To see it, imagine that such a flat space Lagrangian
four--form, say $\mathfrak{L}_{\eta_1}$, exists. Let
$\tilde{\mathfrak{L}}_{\eta_1}$ denote the result of application of MCP
to $\mathfrak{L}_{\eta_1}$. If $\tilde{\mathfrak{L}}_{\eta_1}$ differs
form (\ref{GNC}) by a differential, then the flat space limits of
these two four--forms (obtained for the flat Minkowski metric and
vanishing torsion) ought to differ by a differential as well. But for
(\ref{GNC}) this limit would be just $\mathfrak{L}_{F0}$. Hence,
we would have
\be
\mathcal{L}_{\eta_1}=\mathcal{L}_{F0}+\partial_{\mu}V^{\mu}
\ee
for some vector field $V$.  
But then application of MCP would yield the relation
\be
\tilde{\mathfrak{L}}_{\eta_1}=\tilde{\mathfrak{L}}_{F0}+\stackrel{\circ}{\nabla}_{\mu}V^{\mu}\epsilon-
T_{\mu}V^{\mu}\epsilon=\mathfrak{L}_{F,nm}+\stackrel{\circ}{\nabla}_{\mu}V^{\mu}\epsilon-\left[{\l{\frac{1}{8}+\eta_1}\r
    S_aJ_{(A)}^a}+T_aV^a\right]\epsilon \ .
\ee
Although the second component in the final expression is a
differential, the third one is not (for generic torsion) which
contradicts our assumption.

Hence, we see that (\ref{GNC}) represents a truly non--minimal
coupling. It is a part of a family of couplings
\be\label{NMSh}
\mathfrak{L}_{\frac{1}{2},non-min}={\stackrel{\circ}{\tilde{\mathfrak{L}}}_{F0}}+
\eta_1\,S_aJ^a_{(A)}\,\epsilon+\eta_2\,T_aJ^a_{(V)}\,\epsilon \ ,
\ee
discussed in \cite{Sh}. The second part corresponding to $\eta_2$ is
not really non--minimal, as it arises as a result of the application of
MCP to the flat space Lagrangian constructed from (\ref{LF0}) via
(\ref{Lch}) with $V$ being given by (\ref{Vab}) for $a=-\eta_2$,
$b=0$. It is interesting to note that the truly non--minimal parameter
$\eta_1$ is exactly what is needed to render the appropriately defined
quantum field theory in curved space with torsion renormalizable
\cite{BS1,BS2}. This is why the axial part of torsion tensor $S^a$ was
claimed the most phenomenologically relevant and the trace part of
(\ref{NMSh}) was not even considered in some further developments \cite{BSV}.

\section{Einstein--Cartan gravity with Holst term}\label{ECgravity}

\subsection{General formalism and field equations}\label{GF}

The total Lagrangian four--form of the theory is 
\be
\mathfrak{L}=\mathfrak{L}_G+\mathfrak{L}_{hol}+\mathfrak{L}_m \ ,
\ee
where 
\be
\mathfrak{L}_G=-\frac{1}{4k}\epsilon_{abcd}e^a\wedge e^b\wedge \Omega^{cd} \ , \qquad
\mathfrak{L}_{hol}=\frac{1}{2k\beta}e^a\wedge e^b\wedge \Omega_{ab} 
\ee
and $\mathfrak{L}_m$ represents the matter
part. Here $k=8\pi G$, where $G$ is a gravitational constant, $\beta$
is an Immirzi parameter whose relevance for the quantum theory of
gravity was sketched in the introduction, $e^a=e^a_{\mu}\d x^{\mu}$ is an
orthonormal cotetrad, ${\omega^a}_b={\Gamma^a}_{bc}e^c$ are connection one--forms
(spin connection) obeying antisymmetry condition
$\omega_{ab}=-\omega_{ba}$ and
${\Omega^a}_b:=\d{\omega^a}_b+{\omega^a}_c\wedge{\omega^c}_b=\frac{1}{2}{R^a}_{bcd}e^c\wedge
e^d$ are the
curvature two--forms. The connection coefficients ${\Gamma^a}_{bc}$
are related to the metric connection $\nabla$ on the RC manifold by
$\nabla_{\tilde{e}_c}\tilde{e}_b={\Gamma^a}_{bc}\,\tilde{e}_a$, where
$\tilde{e}_a=e^{\mu}_a\partial_{\mu}$ is an orthonormal tetrad
(a basis of vector fields which is dual to one--form field basis
$e^a$). Variation is given by
\[
\delta{\mathfrak{L}}=\delta e^a\wedge
\l{\frac{\delta\mathfrak{L}_G}{\delta e^a}+\frac{\delta\mathfrak{L}_{hol}}{\delta e^a}+\frac{\delta\mathfrak{L}_m}{\delta e^a}}\r+
\delta\omega^{ab}\wedge
\l{\frac{\delta\mathfrak{L}_G}{\delta\omega^{ab}}+\frac{\delta\mathfrak{L}_{hol}}{\delta\omega^{ab}}+\frac{\delta\mathfrak{L}_m}{\delta\omega^{ab}}}\r+
\delta\phi^A\wedge\frac{\delta\mathfrak{L}_m}{\delta\phi^A} \ ,
\]
$\phi^A$ representing matter fields (we used the independence of
$\mathfrak{L}_G$ and $\mathfrak{L}_{hol}$ on $\phi^A$). Explicitly,
\be
\ba
&\frac{\delta\mathfrak{L}_G}{\delta e^a}=-\frac{1}{2k}\epsilon_{abcd}
e^b\wedge\Omega^{cd} \ , \quad 
\frac{\delta\mathfrak{L}_G}{\delta\omega_{ab}}=-\frac{1}{2k}{\epsilon_{cd}}^{ab}Q^c\wedge
  e^d \ , \\
&\frac{\delta\mathfrak{L}_{hol}}{\delta e^a}=\frac{1}{k\beta}e^b\wedge\Omega_{ab}=\frac{1}{k\beta}DQ_a \ , \quad 
\frac{\delta\mathfrak{L}_{hol}}{\delta\omega_{ab}}=\frac{1}{2k\beta}D\l{e^a\wedge
e^b}\r=\frac{1}{k\beta}Q^{[a}\wedge e^{b]} \ ,
\ea
\ee
where $Q^a:=De^a=\frac{1}{2}{T^a}_{bc}e^b\wedge e^c$ is a torsion
two--form whose components in a tetrad basis we are denoting by
${T^a}_{bc}$. The resulting field equations are 
\be\label{FEq}
\left. 
\begin{array}{cccc}
\dfrac{\delta\mathfrak{L}_G}{\delta e^a}+\dfrac{\delta\mathfrak{L}_{hol}}{\delta e^a}+\dfrac{\delta\mathfrak{L}_m}{\delta e^a}=0 \hfill & \quad \Leftrightarrow \quad &
{G^a}_b:={R^a}_b-\frac{1}{2}R\delta^a_b=k\,{t_b}^a+\dfrac{1}{2\beta}\epsilon^{acde}R_{bcde} \hfill \\
  &   &   \\
\dfrac{\delta\mathfrak{L}_G}{\delta\omega^{ab}}+\dfrac{\delta\mathfrak{L}_{hol}}{\delta\omega^{ab}}+\dfrac{\delta\mathfrak{L}_m}{\delta\omega^{ab}}=0 \hfill & \quad \Leftrightarrow \quad &
T^{cab}-T^a\eta^{bc}+T^b\eta^{ac}=\dfrac{k\beta^2}{1+\beta^2}\l{S^{abc}-\dfrac{1}{2\beta}{\epsilon^{ab}}_{de}S^{dec}}\r
\hfill \\
  &   &   \\
\dfrac{\delta\mathfrak{L}_m}{\delta\phi^A}=0 \hfill &  & \\
\end{array} 
\right.
\ee
where ${R^a}_b:=\eta^{ac}{R^d}_{cdb}$, $R:={R^a}_a$, $T^a:={T^{ba}}_
b$ and the dynamical definitions of energy--momentum and spin density tensors on
Riemann--Cartan space are
\be\label{ts}
{t_a}^b\epsilon:=\dfrac{\delta\mathfrak{L}_m}{\delta e^a}\wedge e^b \ , \qquad
S^{abc}\l{\star e_c}\r:=2\dfrac{\delta\mathfrak{L}_m}{\delta\omega_{ab}} \ .
\ee

\subsection{Effective action}\label{effective}

If the spin density tensor $S^{abc}$ does not depend on the
connection, which is the case for fermions modelled by the Dirac
Lagrangian, the second equation of (\ref{FEq}) represents an
invertible\footnote{Note that this is not a generic feature of EC
  theory and may not be true if the spin density tensor depended on
  torsion. For example, in the case of Proca field invertibility
  breaks down for some values of the field, which gives rise to the
  notion of {\it torsion singularities} \cite{HHKN}.} algebraic relation between the components of the torsion
tensor and the spin density tensor. We can see from (\ref{FEq}) that
this important feature of EC theory remains true after the addition of
the Holst term (recall that real values of the Immirzi parameter are
considered in this paper). This makes the torsion vanish
wherever the  distribution of matter vanishes (the {\it torsion waves} does
not exist). Then the connection becomes the Levi--Civita one,
determined by the metric, and the first equation of (\ref{FEq})
reduces to the usual vacuum Einstein equation (note that the $\beta$
dependent term of this equation vanishes then on account of Bianchi
identity). This is a desirable feature of both the theories (with and
without the Holst term), as it renders them compatible with all the
experimental tests of GR that are based on vacuum solutions.

Wherever the spin density does not vanish, a nonzero torsion must
appear. This may have significance either for the classical theory
of self--graviting matter (star formation, singularity theorems etc.) or
for the semi--classical description of quantum fields. In the latter
case, the EC theory is believed to differ from GR by the presence of
gravity--induced point fermion interaction. The character of this
interaction for different Dirac Lagrangians was studied in
\cite{Kazm}. Here we will enquire whether this character changes after the addition of Holst term to the action.

In order to investigate the physics emerging from the theory for the
space--time metric approaching the flat Minkowski's one and to compare
the predictions of GR, EC theory and Holst--modified EC theory, it is
extremely useful to 
express the torsion through matter fields by means of second equation
of (\ref{FEq}) and to insert the result back to the initial action. In this way an {\it effective
action} is obtained, which does not depend on torsion anymore. We
shall now derive this action for Holst--modified gravity with fermions. Let us define the {\it contortion one--forms}
\[
{K^a}_b={K^a}_{bc}\,e^c:={\omega^a}_b-\stackrel{\circ}{\omega}{{^a}_b}
\]
(objects with $\circ$ above will always denote torsion--free objects,
related to LC connection). The curvature two--form decomposition
\[
{\Omega^a}_b=\stackrel{\circ}{\Omega}{^a}_b+{\stackrel{\circ}{D}\,}{K^a}_b+{K^a}_c\wedge{K^c}_b
\]
results in
\beq
\ba
&\mathfrak{L}_G:=-\frac{1}{4k}\,\epsilon_{abcd}\,e^a\wedge e^b\wedge{\Omega^{cd}}
={\stackrel{\circ}{\mathfrak{L}}_G}-\frac{1}{4k}\epsilon_{abcd}\,e^a\wedge e^b\wedge{K^c}_e\wedge{K^{ed}}
-\frac{1}{4k}{\,\stackrel{\circ}{D}}\l{\,\epsilon_{abcd}\,e^a\wedge e^b\wedge{K^{cd}}\,}\r \ ,
\ea
\eeq
where ${\stackrel{\circ}{D}}\epsilon_{abcd}=0$ was used.
Here $k=8 \pi G$, where
$G$ is a gravitational constant. Since all Lorentz indexes in the
last term are contracted, ${\stackrel{\circ}{D}}$ acts like a usual differential. 
Similarly, for Holst term we get
\be
\mathfrak{L}_{Hol}:=\frac{1}{2k\beta}e^a\wedge e^b\wedge \Omega_{ab}=
\frac{1}{2k\beta}e^a\wedge e^b\wedge K_{ac}\wedge{K^c}_b+\frac{1}{2k\beta}{\,\stackrel{\circ}{D}}\l{e^a\wedge e^b\wedge K_{ab}}\r \ ,
\ee
which yields the combined result
\be
\mathfrak{L}_G+\mathfrak{L}_{Hol}={\stackrel{\circ}{\mathfrak{L}}_G}+
\frac{1}{2k\beta}e^a\wedge e^b\wedge\l{K_{ac}\wedge{K^c}_b-\frac{\beta}{2}\epsilon_{abcd}{K^c}_e\wedge K^{ed}}\r 
+\d (\dots )
\ee
(the last term is a differential and its particular form will not be
needed).
Using the relation between components of contortion and torsion
tensors
\be
K_{abc}=\frac{1}{2}(T_{cab}+T_{bac}-T_{abc})
\ee
and decomposing torsion into its irreducible parts
\be
\ba
&T_{abc}=\frac{1}{3}(\eta_{ac}T_b-\eta_{ab}T_c)+\frac{1}{6}\epsilon_{abcd}S^d+q_{abc}
\ , \\
&T_a:={T^b}_{ab} \ , \ S_a:=\epsilon_{abcd}T^{bcd} \ ,
\ea
\ee
we can finally obtain 
\be
\mathfrak{L}_G+\mathfrak{L}_{Hol}={\stackrel{\circ}{\mathfrak{L}}_G}+\frac{1}{2k}
\l{\frac{2}{3}T_aT^a-\frac{1}{3\beta}T_aS^a-\frac{1}{24}S_aS^a-\frac{1}{2}q_{abc}q^{abc}-
\frac{1}{4\beta}\epsilon^{abcd}q_{eab}{q^e}_{cd}}\r\,\epsilon+\d (\dots) \ .
\ee
Similarily, the Dirac Lagrangian four--form decomposes
according to (\ref{LFdecomp}). The addition of a divergence of a vector field $V$ to the flat space
Lagrangian results in one more term (\ref{TV}). Ultimately, we have the following four--form on
RC space representing gravity with Holst term and fermions
\be\label{Ltot}
\ba
\mathfrak{L}={\stackrel{\circ}{\mathfrak{L}}_G}+{\stackrel{\circ}{\tilde{\mathfrak{L}}}_{F0}}
+\frac{1}{2k}\l{\frac{2}{3}T_aT^a-\frac{1}{3\beta}T_aS^a-\frac{1}{24}S_aS^a-\frac{1}{2}q_{abc}q^{abc}-
\frac{1}{4\beta}\epsilon^{abcd}q_{eab}{q^e}_{cd}-\frac{k}{4}S_aJ^a_{(A)}-2kT_aV^a}\r\,\epsilon \ .
\ea
\ee
The total differential has been omitted in the final formula. Variation of the resulting action with respect to $T_a$, $S_a$ and $q_{abc}$
yields the equations
\be\label{TSV}
\ba
T^a=\frac{3k\beta}{4(1+\beta^2)}\l{2\beta V^a-J^a_{(A)}}\r \ , \qquad
S^a=\frac{-3k\beta}{1+\beta^2}\l{2 V^a+\beta J^a_{(A)}}\r \ , \qquad
q_{abc}=0 \ .
\ea
\ee
Inserting these results into (\ref{Ltot}) we finally get the effective
Lagrangian four--form
\be
\mathfrak{L}_{eff}={\stackrel{\circ}{\mathfrak{L}}_G}+{\stackrel{\circ}{\tilde{\mathfrak{L}}}_{F0}}+
\frac{3k\beta^2}{16(1+\beta^2)}\l{J^{(A)}_aJ_{(A)}^a-4V_aV^a+\frac{4}{\beta}J^{(A)}_aV^a}\r\,\epsilon \ .
\ee
By taking the limit ${\beta \to \infty}$ we can recover the effective
Lagrangian for the usual EC theory
\be\label{ECeff}
\mathfrak{L}_{EC \ eff}={\stackrel{\circ}{\mathfrak{L}}_G}+{\stackrel{\circ}{\tilde{\mathfrak{L}}}_{F0}}+
\frac{3k}{16}\l{J^{(A)}_aJ_{(A)}^a-4V_aV^a}\r\,\epsilon \ ,
\ee
whose physical implications were discussed in \cite{Kazm}. One of the
striking differences between the theory with and without Holst term
concerns the consequences of the requirement of parity invariance of
the finall theory. If Holst term is not present, this condition
appears to be equivalent to the requirement of parity invariance of
the flat space Lagrangian density for fermions. After it is imposed,
we are left with the one--parameter family of theories, parameterized by
$a$ (parameter $b$ has to be equal to zero). It seems difficult to
provide theoretical arguments which would allow for further restriction
of the remaining freedom, except for
some speculations based on the resulting form of the spin density tensor
\cite{Kazm}. Things look different if we adopt Holst--modified
Lagrangian as representing gravitational field. 
Here, the requirement of parity invariance of the
final theory fixes things uniquely. To see that this is the case, note
that the second equation in (\ref{TSV}) is parity invariant if and
only if $V$ is an axial vector. But then, the first equation in (\ref{TSV}) is parity
invariant only if $V^a=\frac{1}{2\beta}J_{(A)}^a$. This choice
corresponds to what was done in \cite{Merc} (the whole family of
couplings considered in \cite{Merc} can be obtained by taking
$V^a=\frac{1}{2\alpha}J_{(A)}^a$). Hence, it is interesting to note
that Holst modification together with the symmetry requirements provides
us with uniqueness, which seemed to be impossible on the grounds of
the standard EC theory.

Note that this result does not decide the question of measurability of
the Immirzi
parameter. After appropriate experiments are performed, whether it is possible to decide about the physical relevance
of Holst term or not will depend on the particular outcomes. To see it,
note that although the EC theory with $V=0$ is
indeed equivalent to the Holst--modified theory with
$V=\frac{1}{2\beta}J_{(A)}$, the EC theory with $V\not=0$ is NOT
equivalent to the Holst--modified theory for any choice of $V$ in the
latter one, as can be easily seen from equations (\ref{TSV}) and
their limiting forms for ${\beta \to \infty}$. Let us imagine that we can actually measure the torsion
itself and after a series of clever experiments we have
established that the torsion trace vector $T^a$ assumes a certain
non--zero value. Then we can try to determine $S^a$. If EC theory
without Holst term is valid, we should get $S^a=-3kJ_{(A)}^a$. If
Holst--modified theory is appropriate, we ought to have
$S^a=-3kJ_{(A)}^a-\frac{4}{\beta}T^a$ instead. In the latter case, the
result would provide an information about the value of Immirzi parameter.

\subsection{Is it possible to distinguish between GR, EC and Holst--modified EC theory by measuring the strength of interactions?}\label{distingtion}

In the previous section we have argued that it is possible that we
could choose between EC and Holst--modified EC theory by measuring
torsion. Here we shall consider the more realistic possibility, based
on measurements of strength of point interactions between fermions. In the
limit of vanishing Riemannian curvature, the effective fermionic
Lagrangian density for all theories under consideration assumes the form
\be\label{flateff}
\mathcal{L}_{F \ eff}={\mathcal{L}_{F0}}+
C_{AA} \ J^{(A)}_aJ_{(A)}^a+C_{AV} \ J^{(A)}_aJ_{(V)}^a+C_{VV} \ J^{(V)}_aJ_{(V)}^a \ .
\ee
For GR, all coupling constants vanish. For EC theory we have
\be\label{ECc}
C_{AA}=\frac{3k}{16}(1-4b^2) \ , \qquad C_{AV}=-\frac{3k}{2}ab \ ,
\qquad C_{VV}=-\frac{3k}{4}a^2 \ , 
\ee
whereas for Holst--modified EC theory
\be\label{ECHc}
C_{AA}=\frac{3k\beta}{16(1+\beta^2)}\left[{4b+\beta(1-4b^2)}\right] \ , \qquad 
C_{AV}=\frac{3k\beta}{4(1+\beta^2)}a(1-2\beta b) \ , \qquad C_{VV}=-\frac{3k\beta^2}{4(1+\beta^2)}a^2 \ . 
\ee
As long as we get experimental values of all coupling constants indistinguishable
from zero, we are not able to say which of the three theories of
gravitation is correct. Measuring a non--zero value of at least one of
them would provide an argument against standard GR. Of course, for any
values of the coupling constants, one could adopt from the beginning (\ref{flateff}) itself as representing fermionic field in
Minkowski space--time and use the torsionless approach of standard GR
to include gravity. The presence of a point fermion interaction does
not really contradict GR. However, on the grounds of EC and Holst--modified EC
theory, the interaction terms arise naturally as a necessary
consequence of the relation between torsion and matter, which was explained
in Subsection \ref{effective}. In this paper, we aim to treat all
the theories in the most natural manner, adopting the simplest Dirac
theory of fermions in flat space as a starting point in each case. Then,
the most traditional method of minimal coupling is used to incorporate
gravity and the results are compared.

According to this standpoint, such a non--zero value of a coupling
constant would discredit GR. It could
either agree with both remaining theories, which would be the case if
the sets of equations (\ref{ECc}) and (\ref{ECHc}) had a solution for
measured values of the coupling constants, or contradict both of
them. It appears that we cannot get a result which would agree with EC
theory and disagree with Holst--modified theory, or conversely. For
given values of the $C$--constants, either both (\ref{ECc}) and (\ref{ECHc}) have a
solution with respect to $a$ and $b$, or both are inconsistent. To see
this, note that the change of parameters
\be
a\rightarrow -\frac{\beta}{\sqrt{1+\beta^2}}a \ , \qquad
b\rightarrow \frac{1-2\beta b}{2\sqrt{1+\beta^2}}
\ee
in (\ref{ECc}) leads directly to (\ref{ECHc}). 
Hence, as long as we cannot cope with ambiguities resulting from the
freedom of addition of a divergence to the flat Lagrangian, 
it is not possible to discriminate between EC theory and Holst--modified EC theory by measuring the strength of four--fermion
interactions, although it is in principle possible to rule them both
out, or to strengthen their position and discredit GR. 

\section{Conclusions}\label{conc}
The ambiguity of the minimal coupling procedure in the presence of
torsion allows for the reinterpretation of all the non--minimal coupling
procedures considered in the literature concerning Holst--modified
gravity with fermions. They can be viewed as minimal couplings for
appropriate flat space Lagrangians for fermions. There exist genuine
non--minimal couplings which cannot be viewed in this way. 

After some reasonable requirements are imposed on the Lagrangian formulation of
the theory of Dirac field in flat space, the
above--mentioned ambiguity is reduced to the two--parameter freedom in the final theory of Holst--modified EC gravity with
fermions. The resulting family of theories is equivalent to the
combined families considered in \cite{FMT} and
\cite{Merc}. The richer four--parameter family of couplings
introduced in \cite{Alex} consists of the above mentioned
two--parameter one, scaling and the new one--parameter
family. However, the latter is not physically relevant from the
viewpoint adopted in this article, as it corresponds to the flat space
Lagrangian for mass--less fermions which does not allow for consistent
addition of a mass term.

Unlike the standard EC theory, the theory with Holst term becomes
unique under the requirement for it to be parity invariant. This property can be traced back to the difference in the behavior of
standard gravitational action and Holst term under parity
transformation. The resulting unique theory is the one that is usually
considered as EC gravity with fermions -- in fact, the EC gravity with
fermions and without Holst term necessarily contains a one--parameter ambiguity, even if
parity invariance requirement is imposed. This surprising feature of the
Holst--modified theory is not conclusive, since there is no reason to
assume a priori that parity should not be broken by gravitational
interaction. 

As long as we cannot solve the nonuniqueness problem theoretically,
the theories with and without Holst term are indistinguishable on the
effective level. This means that it is impossible to observe the
physical effects of the Immirzi parameter by measuring the strength of
gravity--induced interactions between fermions. However, if we were
able to perform direct measurements of the space--time torsion, the
effects of the Immirzi parameter could be detectable. The final answer
to the question of its measurability would depend on the particular
outcomes of experiments. We wish to stress that all the conclusions resulting
from the analyses of this paper reflect the current state of knowledge
and may intrinsically change if a satisfactory solution to the problem of
nonuniqueness of EC theory is found.

\section*{Acknowledgements}
I would like to thank Wojciech Kami{\'n}ski, Piotr Kosi{\'n}ski, Jerzy Lewandowski and
Ilya Shapiro for helpful comments and Urszula Pawlik for linguistic corrections.
This work was partially supported by the Foundation for Polish Science, Master grant.

\section{Appendix: Notation and conventions}\label{App1}

Throughout the paper we use the units $c=\hbar =1$. The indexes $a,b,\dots$
correspond to an orthonormal tetrad, whereas $\mu,\nu,\dots$ correspond to
a holonomic frame. For inertial frame of flat Minkowski space,
which is both holonomic and orthonormal, we use $\mu,\nu,\dots$.
The metric components in an orthonormal tetrad basis $\tilde{e}_a$ are
 $g\l{\tilde{e}_a,\tilde{e}_b}\r=(\eta_{ab})=diag(1,-1,-1,-1)$. By
 $e^a$ we shall denote an orthonormal cotetrad -- the basis of
 one--form fields which is dual to the tetrad basis, $e^a\l{\tilde{e}_b}\r=\delta^a_b$. Lorenz
 indexes are shifted by $\eta_{ab}$. $\epsilon=e^0\wedge e^1\wedge
e^2\wedge e^3$ denotes the cannonical
volume four--form  whose components in orthonormal tetrad basis obey $\epsilon_{0123}=-\epsilon^{0123}=1$.
The action of a covariant exterior differential $D$ on any $(r,s)$-tensorial
type differential $m$--form 
\beq
{T^{a_1\dots a_r}}_{b_1\dots b_s}=
\frac{1}{m!}{T^{a_1\dots a_r}}_{b_1\dots b_s\mu_1\dots \mu_m}\d x^{\mu_1}\w\dots\w\d x^{\mu_m}
\eeq
is given by 
\beq
D{T^{a_1\dots a_r}}_{b_1\dots b_s}:=\d {T^{a_1\dots a_r}}_{b_1\dots b_s}+
\sum_{i=1}^r{\omega^{a_i}}_c\w {T^{a_1\dots c\dots a_r}}_{b_1\dots b_s}-
\sum_{i=1}^s{\omega^c}_{b_i}\w {T^{a_1\dots a_r}}_{b_1\dots c\dots b_s} \ .
\eeq
The covariant derivative of a Dirac bispinor field is
\be
D\psi=\l{D_a\psi}\r\, e^a:=\d\psi-\frac{i}{2}\omega_{ab}\Sigma^{ab}\psi \ , \quad 
\ov{D\psi}=\l{D\psi}\r^{\dagger}\gamma^0  \ , \quad
\Sigma^{ab}:=\frac{i}{2}[\gamma^a,\gamma^b] \ ,
\ee
where $\gamma^a$ are Dirac matrixes, $\omega_{ab}=-\omega_{ba}$ are
connection one--forms (spin connection). The hodge star action on external products of orthonormal cotetrad
one--forms is given by
\beq
\star e_a=\frac{1}{3!}\epsilon_{abcd}e^b\w e^c\w e^d \ , \quad 
\star \l{e_a\w e_b}\r=\frac{1}{2!}\epsilon_{abcd}e^c\w e^d \ , \quad 
\star \l{e_a\w e_b\w e_c}\r=\epsilon_{abcd}e^d \ ,
\eeq
which by linearity determines the action of $\star$ on any differential
form.

\end{document}